\documentclass[superscriptaddress,amsmath, nofootinbib]{revtex4}   %BASIC
\usepackage{amsfonts}
\usepackage{euscript}
\usepackage{graphicx}
\usepackage[latin1]{inputenc}
\usepackage{amsmath}
\usepackage{amssymb}
\usepackage[latin1]{inputenc}
\usepackage{amsmath}
\usepackage{amssymb}
\usepackage{amsfonts}
\usepackage{graphicx}
\usepackage[latin1]{inputenc}
\usepackage{amsmath}
\usepackage{amssymb}
\begin{document}
%\draft

%\preprint{APS/123-QED}

\title{Comment on "Frame-dragging: meaning, myths, and misconceptions" by L. F. O. Costa and J. Nat\'ario} 
\author{Alexei A. Deriglazov }
\email{alexei.deriglazov@ufjf.edu.br} 
\affiliation{Depto. de Matem\'atica, ICE, Universidade Federal de Juiz de Fora,
MG, Brazil,} \affiliation{Department of Physics, Tomsk State University, Lenin Prospekt 36, 634050, Tomsk, Russia}

\date{\today}% It is always \today, today,

\begin{abstract}
I point out that the authors' interpretation of their calculations differs from the standard interpretation, described in Sect. 84 of Landau-Lifshitz book.  
This casts doubt on the authors' claim that Sagnac effect "{\it arises also in apparatuses which are fixed relative to the distant stars (i.e., to asymptotically inertial frames); in this case one talks  about frame-dragging.}".  
\end{abstract}

\maketitle %\noindent
%%%%%%{\bf DOI:}
%%%%%%%{\bf PACS numbers:} 11.10.Ef, 03.65.Ca \\
%%%%%%%%{\bf Keywords:} Zitterbewegung, Semiclassical Description of Relativistic Spin, Dirac Equation, Theories with Constraints.
%   {Mathisson:1937zz} {Papapetrou:1951pa} {Tulc} {Dixon1964} {Dixon1965} {pirani:1956}

\section{The comment.}

In a recent work \cite{Costa_N_2021}, Costa and Nat\'ario discussed the notion of frame-dragging in stationary ($g_{\mu\nu}({\bf x}$)) rotating ($g_{0i}\ne 0$) gravitational fields.  The interval of such spacetime can be written in $1+3$ block-diagonal form as follows:
\begin{eqnarray}\label{fd.1}
ds^2=c^2\left[\sqrt{-g_{00}}(dt-A_idx^i)\right]^2-dl^2, 
\end{eqnarray}
\begin{eqnarray}\label{fd.1.1}
\quad A_i\equiv -\frac{g_{0i}}{cg_{00}}, \qquad dl^2\equiv h_{ij}dx^idx^j, \qquad h_{ij}\equiv g_{ij}-\frac{g_{0i}g_{0j}}{g_{00}}, \qquad  e^{\Phi}\equiv\sqrt{-g_{00}}. 
\end{eqnarray}
A worldline $x^\mu=(ct, {\bf x})$, where $t\in \mathbb{R}$ while ${\bf x}=\mbox{const}$, is called an observer {\it at rest} (or laboratory observer) \cite{Costa_N_2021}, see the discussion around Eq. (2) in \cite{Costa_N_2021}.

In Sect. 2.1, after mentioning the Sagnac effect in flat spacetime (see Sect. 89 in \cite{Landau_2}), the  authors  state: {\it "In a gravitational field, however, it}  (Sagnac effect) {\it arises also in apparatuses which are fixed relative to the  distant stars (i.e., to asymptotically inertial frames); in this case one talks  about frame-dragging."}. To confirm the claim, Costa and Nat\'ario used the four dimensional geometry of general relativity (\ref{fd.1}), and considered  a thought experiment, that implies the necessity of the following calculation: a photon was emitted by observer at rest at a spatial point $O$, and then propagate along a null path $x^\mu(\lambda)$, with known ${\bf x}(\lambda)$, arriving at the end to the point $O$: ${\bf x}(0)={\bf x}(1)$. Knowing the closed loop $\bf{x}(\lambda)$, the task is to calculate the photons' travel time, measured by $O$.  

To this aim, Costa and Nat\'ario separated the coordinate time $dt$ from Eq. (\ref{fd.1}) with $ds^2=0$, obtaining 
%\begin{eqnarray}\label{fd.3}
$dt=A_idx^i+\frac{dl}{c\sqrt{-g_{00}}}$. 
%\end{eqnarray}
Identifying $dt$ with the time measured by $O$, they concluded that  the travel time is 
\begin{eqnarray}\label{fd.3.1}
t=\int_{C}A_idx^i+\int_{C}\frac{dl}{c\sqrt{-g_{00}}}. 
\end{eqnarray}

However, here the point is that the coordinate time $dt$ in (\ref{fd.1}) does not represent the time measured by the laboratory of observer at rest $O$. The latter should be found using the Landau-Lifshitz prescription, that associates the time interval and spatial distance to the infinitesimally closed events $x^\mu$ and $x^\mu+dx^\mu$. The "true" time\footnote{"True" time is the direct translation of the terminology,  used by Landau and Lifshitz in Russian version of \cite{Landau_2}.} is\footnote{See the discussion around unnumbered equation below the Eq. (84.5), and around Eq. (88.9) in \cite{Landau_2}; or Sect. III A in \cite{AAD_Rec}.}
\begin{eqnarray}\label{fd.4}
dt_p=\sqrt{-g_{00}}(dt-A_idx^i), 
\end{eqnarray}
and is just the square root of first term on r.h.s. of Eq. (\ref{fd.1}).  The spatial distance is equal to $dl$, that is to the square root of second term.  
When the events $x^\mu$ and $x^\mu+dx^\mu$ represent the emission and absorption of a photon, these definitions together with the propagation law $ds^2=0$ imply: ${\bf v}^2\equiv\frac{dl^2}{dt_p^2}=c^2$. That is the constancy of the speed of light in a vacuum is implicit in the Landau-Lifshitz prescription, see also \cite{AAD_Rec}.

Using (\ref{fd.1}) with $ds^2=0$, the Landau-Lifshitz prescription (\ref{fd.4}) implies the travel time 
\begin{eqnarray}\label{fd.5}
t_{p}=\frac1c\int_{C}dl, 
\end{eqnarray}
which is different from (\ref{fd.3.1}). 

All this casts doubt on the authors' claim that the Sagnac effect "{\it arises also in apparatuses which are fixed relative to the  distant stars (i.e., to asymptotically inertial frames); in this case one talks  about frame-dragging.}". 

In conclusion, two comments are in order. \par 
\noindent 
1. Even for non rotating metric ($A_i=0$), the intervals $dt$ and $dt_p$ still different due to the conversion factor $\sqrt{-g_{00}}$. Ignoring this factor leads to a number of erroneous results \cite{Gruder_2017}, see the discussion in \cite{Gron_2018, AAD_Gr}.  \par 
\noindent   
2. In discussing their equations (84.1)-(84.7), Landau and Lifshitz emphasize that their definitions of time and distance do not require any prior synchronization of clocks.  

\section{Note added.}\par 
\noindent {\bf 1.} In their computations, authors of \cite{Costa_N_2021} used $ds^2=0$, and assumed that photon moves along a prescribed finite path $C={\bf x}(\lambda)$. In other words, \underline{by the word "photon/light signal" the authors mean any null worldline} in spacetime (\ref{fd.1}). As I understand, this was emphasized also in their Reply \cite{Costa_N_2021_2}. \par
\noindent {\bf 2.} The quantity $dt_{RE}=-{\bf A}d{\bf x}+e^{-\Phi}dl$, introduced by authors on page 1 of \cite{Costa_N_2021_2}, 
does not zero Eq. (1) of \cite{Costa_N_2021_2}. So, let me reconsider the two-way trip experiment to see, which equations will appear in the authors' notation (${\bf x}_R={\bf x}_E+d{\bf x}$). 

Observer $E$ emits a light flash at position ${\bf x}_{E}$, which is reflected by the mirror $R$ at infinitesimally closed point ${\bf x}_{R}$, returning then to $E$. Let me collect the "experimental dates" of two-way trip experiment as follows:
\begin{eqnarray}\label{fd.12}
t_{pE}=0, \quad {\bf x}_E, \qquad {\bf x}_R, \qquad  2t_{p}>0. 
\end{eqnarray}
$t_{pE}=0$ and $2t_p$ represent the proper time instants of emission and absorption of the photon at ${\bf x}_E$, measured by the clock of observer at rest at ${\bf x}_E$. The task is to restore the four-dimensional coordinates  $x^\mu_E$, $x^\mu_R$ and $x^\mu_S$ of the events $E$, $R$ and  $S$ ($S$ is the event of arriving to the point ${\bf x}_E$) from the known three-dimensional quantities (\ref{fd.12}) measured in the laboratory. 

In the stationary spacetime (\ref{fd.1}), we can identify $x^0_E=ct_{pE}=0$. Then $x^0_S=c\frac{2t_p}{\sqrt{-g_{00}}}$, and 
we have the events
\begin{eqnarray}\label{fd.13}
x^\mu_E=(0, {\bf x}_E), \qquad x^\mu_S=(c\frac{2t_p}{\sqrt{-g_{00}}}, {\bf x}_E) \qquad x^\mu_R=(x^0_R, {\bf x}_R), 
\end{eqnarray}
and we need to restore the time coordinate $x^0_R$ of the event of reflection of the photon at ${\bf x}_R$.  Each of one-way photons obeys the null-line condition, that according to (\ref{fd.1}) reads
\begin{eqnarray}\label{fd.14}
\left[\sqrt{-g_{00}}(dx^0-c{\bf A}d{\bf x})\right]^2=dl^2, 
\end{eqnarray}
and should be specified for the corresponding differences of coordinates. 

Consider the differences\footnote{Null-line equation is invariant under the substitution $dx^\mu\rightarrow -dx^\mu$.} 
\begin{eqnarray}\label{fd.4.1}
dx^\mu_{ER}=x^\mu_R-x^\mu_E=(x^0_R, d{\bf x}),  \qquad dx^\mu_{RS}=x^\mu_R-x^\mu_S=(x^0_R-\frac{2ct_p}{\sqrt{-g_{00}}}, d{\bf x}), \qquad  dx^\mu_{ES}=x^\mu_S-x^\mu_E=(c\frac{2t_p}{\sqrt{-g_{00}}}, {\bf 0}), 
\end{eqnarray}
The definition (\ref{fd.4.1}) implies the identity
%\begin{eqnarray}\label{fd.7}
$dx^\mu_{ES}=dx^\mu_{ER}-dx^\mu_{RS}$. 
%\end{eqnarray}
In particular, the interval of coordinate time for the complete trip $E\rightarrow R\rightarrow S$ is given by the {\it difference} 
\begin{eqnarray}\label{fd.8}
dt_{ES}=dt_{ER}-dt_{RS}=2e^{\phi}dl.
\end{eqnarray}
For the trip $E\rightarrow R$,  Eq. (\ref{fd.14}) reads
\begin{eqnarray}\label{fd.15}
\left[\sqrt{-g_{00}}\left(x^0_R-c{\bf A}d{\bf x}\right)\right]^2=dl^2, 
\end{eqnarray}
while for the trip $R\rightarrow S$
\begin{eqnarray}\label{fd.16}
\left[\sqrt{-g_{00}}\left(x^0_R-c{\bf A}d{\bf x}-\frac{2ct_p}{\sqrt{-g_{00}}}\right)\right]^2=dl^2. 
\end{eqnarray}
Comparing the two equations, we have
\begin{eqnarray}\label{fd.17}
\left[\sqrt{-g_{00}}\left(x^0_R-c{\bf A}d{\bf x}\right)\right]^2=\left[\sqrt{-g_{00}}\left(x^0_R-c{\bf A}d{\bf x}-\frac{2ct_p}{\sqrt{-g_{00}}}\right)\right]^2. 
\end{eqnarray}
Due to the condition $2t_p>0$, the only solution of this equation is (this is Eq. (\ref{fd.4}) in other notation)
\begin{eqnarray}\label{fd.18}
x^0_R=\frac{ct_p}{\sqrt{-g_{00}}}+c{\bf A}d{\bf x}.
\end{eqnarray}
Substitution of the obtained $x^0_R$ back into the equation (\ref{fd.15}) or (\ref{fd.16}) gives
\begin{eqnarray}\label{fd.18.1}
(ct_p)^2=dl^2,
\end{eqnarray}
so the coordinate $x^0_R$  can be presented also in the form
\begin{eqnarray}\label{fd.18.2}
x^0_R=\frac{dl}{\sqrt{-g_{00}}}+c{\bf A}d{\bf x}.
\end{eqnarray}
Using this expression to 
compute the difference $dx^0_{ER}=x^0_R-x^0_E$ of coordinate times for the trip $E\rightarrow R$, we obtain 
\begin{eqnarray}\label{fd.5}
dx^0_{ER}=c{\bf A}d{\bf x}+\frac{dl}{\sqrt{-g_{00}}},
\end{eqnarray}
while  $dx^0_{RS}=x^0_R-x^0_S$ for the trip $R\rightarrow S$ gives  
\begin{eqnarray}\label{fd.6}
dx^0_{RS}=c{\bf A}d{\bf x}-\frac{dl}{\sqrt{-g_{00}}}.
\end{eqnarray}
They coincide with Eqs. (84.5) of the Landau-Lifshitz book. Observe that the term with $dl$, but not $c{\bf A}d{\bf x}$, changes sign with an inversion of direction in these notations. 

Comments. \par 

\noindent (I) $t_p$ is one-half of the total proper time $2t_p$ of the two-way trip. According to Landau-Lifshitz,  from Eqs. (\ref{fd.8}) and (\ref{fd.18.1}) the observer $E$ can interpret $t_p$ as the time of one-way trip. Assuming this,  the one-way photons passed equal distances at equal time $t_p$, with the speed equal to the speed of light $c$. By this way, with a pair of events separated by $(dx^0=cdt, d{\bf x})$, they associated the time interval $dt_p$, that should be computed according to Eq. (\ref{fd.4}). \par

\noindent (II) To give an interpretation of $t_p$ in terms of the four-dimensional geometry, Landau and Lifshitz introduced the event $x^\mu_I=(\frac{ct_p}{\sqrt{-g_{00}}}, {\bf x}_E)$ on the worldline of observer $E$, that he will consider as the event simultaneous with $x^\mu_R$. The event $x^\mu_I$ can be used for synchronization of the clocks of observers ${\bf x}_E$ and ${\bf x}_R$. This was used by Landau and Lifshitz in their book \cite{Landau_2} to discuss the Sagnac effect in Minkowski space, see Sect. 89 entitled "Rotation". 

\begin{acknowledgments}
The work has been supported by the Brazilian foundation CNPq (Conselho Nacional de
Desenvolvimento Cient\'ifico e Tecnol\'ogico - Brasil),  and by Tomsk State University Competitiveness Improvement
Program.
%by the grant from The Tomsk State University D. I. Mendeleev Foundation Programm.
\end{acknowledgments}

\end{document}